# On The Active Input-Output Feedback Linearization of Single-Link Flexible Joint Manipulator: An Extended State Observer Approach

Wameedh Riyadh Abdul-Adheem
Electrical Engineering Department
College of Engineering, Baghdad University
Baghdad, Iraq
Wameedh.r@coeng.uobaghdad.edu.iq

Ibraheem Kasim Ibraheem
Electrical Engineering Department
College of Engineering, Baghdad University
Baghdad, Iraq
ibraheemki@coeng.uobaghdad.edu.iq

*Abstract*—**Traditional input-output feedback linearization (IOFL) is an essential part of nonlinear control theory and a valuable tool in solving class of problems possessing certain constraints. It requires full knowledge of system dynamics and assumes no disturbance at the input channel and no system's uncertainties. In this paper, an Active Input-Output Feedback Linearization (AIOFL) technique based on extended state observer which is the core part of the Active Disturbance Rejection Control (ADRC) paradigm is proposed to design a feedback linearization control law. This control law transforms the system into a chain of integrators up to the relative degree of the system. The proposed AIOFL simultaneously cancels the generalized disturbances (exogenous disturbance and internal uncertainties) and delivers the estimated system's states to the nonlinear state error feedback of the ADRC. Verification of the outcomes has been achieved by applying the proposed technique on the ADRC of Flexible Joint Single Link Manipulator(SLFJM). The results showed the effectiveness of the proposed tool.**

I. INTRODUCTION

There are numerous classes of nonlinear models. Given the following one,

$$\begin{cases} \dot{x} = f(x) + g(x)u \\ y = h(x) \end{cases} \quad (1)$$

Where $x = (x_1 \ x_2 \ \cdots \ x_n)^T \in \mathbb{R}^n$ is the state vector, $u \in \mathbb{R}$ is the control input and $y \in \mathbb{R}$, is the system output. The functions $f, g,$ and $h$ are sufficiently smooth in a domain $D \subset \mathbb{R}^n$. The mappings $f: D \to \mathbb{R}^n$ and $g: D \to \mathbb{R}^n$ are called vector fields on $D$. Consider the Jacobian linearization of the system (1) about the equilibrium point $(x_0, y_0, u_0)$,

$$\begin{cases} \dot{x} = \left[\dfrac{\partial f(x_0)}{\partial x} + \dfrac{\partial g(x_0)}{\partial x} u_0\right](x - x_0) + g(x_0)(u - u_0) \\ y - y_0 = \dfrac{\partial h(x_0)}{\partial x}(x - x_0) \end{cases}$$

It is worthy to observe that the nonlinear system is accurately represented by the Jacobian model only at the equilibrium point $(x_0, y_0, u_0)$. Consequently, any control policy built on the linearized model may produce inacceptable performance at other operating points. Feedback linearization (FL) is another class of nonlinear control methods that can yields linear models that is a precise depiction of the fundamental nonlinear model among a wide set of the equilibrium points [1]. Simply, FL is a technique, which eliminates all the nonlinearities with the result that the nonlinear dynamical system is represented by a chain of integrators. FL can be applied in three steps, the first step is transforming the nonlinear system into a linearized model, this is achieved through appropriate nonlinear change of variables. After this stage, the equations of the system are linear but with the cost of a nonlinear control law($u$) which converts the system into a chain of integrators up to the relative degree of the system with linear control law($v$). The second step is applying one of traditional linear control methods such as state-feedback, PID control, etc., to control the linearized model. The third step is the stability investigation of the internal dynamics [2]. FL has been applied in recent years in various research and industrial fields, for example, in the control of induction motors [3], spacecraft models that includes reaction wheel configuration [4], surface permanent-magnet-synchronous generator (SPMSG) [5]. Further applications include, maximum power point tracking (MPPT) technique to achieve the desired performance under sudden irradiation drops, set-point changes, and load disturbances [6], Adaptive Input-Output Feedback Linearization (AIOFL) to damp the low-frequency oscillations in power systems [7] and for the reduction of torque ripple of a brushless DC motors [8]. Finally, in robust nonlinear controller design for the voltage-source converters of high-voltage, direct current transmission link using IOFL and sliding-mode control approach [9].

This paper proposes a method for Input-Output Feedback Linearization (IOFL) in an active manner, namely, the AIOFL, in which the nonlinearities, model uncertainties, and external disturbance are excellently estimated and canceled, resulting the nonlinear system is reduced into a chain of integrators up to the relative degree of the system. The key points of the proposed method are, AIOFL requires only the relative degree of the nonlinear system in contrast to conventional feedback linearization, which requires complete knowledge of the nonlinear system to design the linearizing control law. The second key point is that there is no need to do any diffeomorphism transformation. Finally, on contrast to conventional IOFL, the proposed AIOFL is highly immune against uncertainties and external disturbances.

This paper is organized as follows. In section II a brief introduction to ADRC is presented. Background and problem statement are introduced in section III. In section IV Active Input-Output Feedback Linearization is discussed with detailed proofs. In section V, the single link flexible joint manipulator is presented as a guideway example for the proposed method. Finally, the conclusions are drawn in section VI.

## II. EXTENDED STATE OBSERVER

Extended State Observer (ESO) is the central part of a recent robust control paradigm, the so-called ADRC. It is an advanced robust control strategy, which works by augmenting the mathematical model of the nonlinear dynamical system with an additional virtual state. This virtual state describes all the unwanted dynamics, uncertainties, and exogenous disturbances and named the "*generalized disturbance*" or "*total disturbance*". This virtual state together with the states of the dynamic system are observed in real-time fashion using the ESO. It performs direct and active prediction and cancelation to the total disturbance by feeding back the estimated generalized disturbance into the input channel after simple manipulation. With ADRC, controlling a complex time-varying nonlinear system is transformed into a simple and linearized process. The superiority that makes it such a successful robust control tool is that it is an error-driven technique, rather than model-based control law. Mainly, ADRC consists of an ESO, a tracking differentiator (TD), and a nonlinear state error combination (NLSEF) as illustrated in Fig. 1 [10]–[12]. Where $r \in \mathbb{R}$ is the reference input, $(r_1 \ r_2 \ \ldots \ r)^T \in \mathbb{R}^n$ is the transient profile, $v \in \mathbb{R}$ is the control input for the linearized model, $(\hat{x}_1 \ \hat{x}_2 \ \ldots \ \hat{x}_{n+1})^T \in \mathbb{R}^{n+1}$ is the augmented estimated vector which comprises the plant's states $\hat{x}_1, \ldots, \hat{x}_n$ and the estimated generalized disturbance $\hat{x}_{n+1}$, which are produced by the extended state observer and $b_0$ is the input gain.

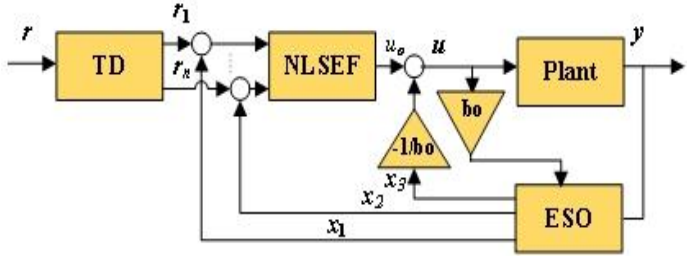

Fig. 1 Structure of ADRC.

*Statement of contribution*. The contribution of this paper is developing the AIOFL technique based on the ESO. We use the ESO not just as an estimator for the total disturbance ($\hat{x}_{n+1}$) which later on will be rejected as shown in Fig.1, but, also as a linearization tool. The nonlinear sliding mode ESO developed in our previous work [12] will be used in this work in addition to the linear ESO (LESO). The advantage of this technique is that it transforms any nonlinear uncertain system with exogenous disturbance into a chain of integrators up to the relative degree of the system. The proposed AIOFL method is effective due its simplicity. For linearization, the only required information is the relative order of the system.

## III. BACKGROUND AND PROBLEM STATEMENT

To perform IOFL, conditions have to be derived and stated which allow us to do the transformation to the nonlinear system such that the input-output map is linear. Given $\dot{y}$ as,

$$\dot{y} = \frac{\partial h(x)}{\partial x}\dot{x} = \frac{\partial h}{\partial x}[f(x) + g(x)u] = L_f h(x) + L_g h(x)u$$

Where $L_f h(x) = \frac{\partial h}{\partial x} f(x)$ is called the *Lie Derivative* of $h(x)$ with respect to $f$. If $L_g h(x) = 0$, then $\dot{y} = L_f h(x)$ is independent of $u$. The second derivative of $y$, denoted by $\ddot{y}$ is given by

$$\ddot{y} = \frac{\partial L_f h}{\partial x}\dot{x} = \frac{\partial L_f h}{\partial x}[f(x) + g(x)u] = L_f^2 h(x) + L_g L_f h(x)u$$

Once again, if $L_g L_f h(x) = 0$, then $\ddot{y} = L_f^2 h(x)$, is also independent of $u$. Repeating this process with $h(x)$, one gets,

$$\begin{cases} L_g L_f^{i-1} h(x) = 0 & i = 1,2,\ldots,\rho-1 \\ L_g L_f^{\rho-1} h(x) \neq 0 \end{cases} \quad (2)$$

It can be seen that $u$ is not included in $y, \dot{y}, \ldots, y^{(\rho-1)}$ but $y^{(\rho)}$ with a nonzero coefficient, $y^{(\rho)} = L_f^\rho h(x) + L_g L_f^{\rho-1} h(x)u$. The control signal $u = \frac{1}{L_g L_f^{\rho-1} h(x)}\left[-L_f^\rho h(x) + v\right]$ reduces the input-output map to $y^{(\rho)} = v$. The system is obviously input-output feedback linearizable, *i.e.* the nonlinear system (1) is represented by a chain of integrators, where $\rho$ is denoted as the *relative degree* of the nonlinear system. Now let

$$z = T(x) = \begin{pmatrix} \phi_1(x) \\ \vdots \\ \phi_{n-\rho}(x) \\ --- \\ h(x) \\ \vdots \\ L_f^{\rho-1} h(x) \end{pmatrix} \stackrel{\text{def}}{=} \begin{pmatrix} \phi(x) \\ --- \\ \psi(x) \end{pmatrix} \stackrel{\text{def}}{=} \begin{pmatrix} \eta \\ -- \\ \xi \end{pmatrix} \quad (3)$$

Where $\phi_1(x)$ to $\phi_{n-\rho}(x)$ are chosen such that $\frac{\partial \phi_i(x)}{\partial x} g(x) = 0$ for $i \in \{1.2.\ldots.n-\rho\} \ \forall \ x \in D$. This condition ensures that when the following equation is calculated $\dot{\eta} = \frac{\partial \phi(x)}{\partial x}[f(x) + g(x)u] = f_0(\eta.\xi) + g_0(\eta.\xi)u$, the term $u$ cancels out. It is now easy to verify that $z = T(x)$ transforms the system into normal form denoted as

$$\begin{cases} \dot{\eta} = f_0(\eta.\xi) \\ \dot{\xi}_i = \xi_{i+1} \ i \in \{1.2.\ldots,\rho-1\} \\ \dot{\xi}_\rho = \alpha(x) + \beta(x)u \\ y = \xi_1 \end{cases}$$

Where $\alpha(x) = L_f^\rho h(x)$ and $\beta(x) = L_g L_f^{\rho-1} h(x)$. The internal dynamics are described by $\dot{\eta} = f_0(\eta, \xi)$. The zero-dynamics of the system is stated as $\dot{\eta} = f_0(\eta, \xi)$ with $\xi = 0$ (*i.e.*, $\dot{\eta} = f_0(\eta, 0)$).

The system is called *minimum phase* if the zero-dynamics of the system are (globally) asymptotically stable.

### IV. ACTIVE INPUT-OUTPUT FEEDBACK LINEARIZATION

Consider the nonlinear SISO system given as

$$y^{(\rho)}(t) = f(y(t), (t). \ldots, y^{(\rho-1)}(t), d(t), t) + \beta(x)u(t)$$

where $y^{(i)}(t)$ indicates $i^{\text{th}}$ derivative of $y$ (the output), and $d$ and $u$ represent the disturbance and the input, respectively. Many classes of nonlinear systems can be represented in this notation, e.g., time-varying or time-invariant systems, nonlinear or linear systems. For simpler representation and without causing any ambiguity, the time variable will be omitted from the equations. Assuming $\xi_1 = y, \xi_2 = \dot{y}. \ldots, \xi_\rho = y^{(\rho-1)}$, one gets

$$\begin{cases} \dot{\xi}_i = \xi_{i+1} \; i \in \{1,2. \ldots, \rho-1\} \\ \dot{\xi}_\rho = f(\xi_1, \xi_2, \ldots, \xi_\rho, w, t) + (\beta(x) - b_0)u + b_0 u \end{cases} \quad (4)$$

Augmenting the system with additional state, $\xi_{\rho+1} = f + (\beta(x) - b_0)u = f_T \Rightarrow \dot{\xi}_{\rho+1} = \Delta(t) = \dot{f}_T$. The coefficient $b_0$ is a rough approximation of $\beta(x)$ in the plant within a $\pm 50\%$ range [10] and $f_T = f + (\beta(x) - b_0)u$ is the generalized disturbance, which consists all of the unknown external disturbances, uncertainties and internal dynamics. The parameter $b_0$ usually chosen explicitly by the user as a design parameter. The states of the system in (4) together with the generalized disturbance $f_T$ will be estimated by the ESO, given by [13], [14],

$$\begin{cases} \dot{\hat{\xi}}_i = \hat{\xi}_{i+1} + \beta_i(y - \hat{\xi}_1) , i \in \{1,2. \ldots, \rho-1\} \\ \dot{\hat{\xi}}_\rho = \hat{\xi}_{\rho+1} + \beta_\rho(y - \hat{\xi}_1) + b_0 u \\ \dot{\hat{\xi}}_{\rho+1} = \beta_{\rho+1}(y - \hat{\xi}_1) \end{cases} \quad (5)$$

The noteworthy feature of the ESO is that it needs minimum information about the dynamical system, only $\rho$ of the underlying system is needed to the design of the ESO. Severa3l modifications have been developed to expand the basic features of ESO to adapt to a broader class of dynamical systems[13]. In this section, the convergence of the Linear ESO (LESO) is considered. Consider the system (4) with the augmented state $\xi_{\rho+1}$ is given as:

$$\begin{cases} \dot{\xi}_i = \xi_{i+1} \; i \in \{1.2. \ldots, \rho-1\} \\ \dot{\xi}_\rho = \xi_{\rho+1} + b_0 u \\ \dot{\xi}_{\rho+1} = \Delta(t) = \dot{f}_T \end{cases} \quad (6)$$

**Assumption (A1):** The function $f_T$ is continuously differentiable
**Assumption (A2):** There exist a positive constant $M$ such that $|\Delta(t)| \leq M$ for $t \geq 0$.
**Assumption (A3)** [14]**:** There exist constants $\lambda_1$, and $\lambda_2$ and positive definite, continuously differentiable functions $V, W: \mathbb{R}^{n+1} \to \mathbb{R}^+$ such that:

$$\lambda_1 \|y\|^2 \leq V(y) \leq \lambda_2 \|y\|^2 , \; W(y) = \|y\|^2, \quad (7)$$

$$\sum_{i=1}^{\rho} \frac{\partial V_i}{\partial y_i}(y_i - a_i y_1) - \frac{\partial V}{\partial y_{\rho+1}} a_{\rho+1} y_1 \leq -W(y). \quad (8)$$

**Lemma 1.** Consider the candidate Lyapunov functions $V, W: \mathbb{R}^{n+1} \to \mathbb{R}^+$ defined by $V(\eta) = <P\eta, \eta>$, where $\eta \in \mathbb{R}^{\rho+1}$ and $P$ is a symmetric positive definite matrix. Suppose that (7) in Assumption (A3) with $\lambda_1 = \lambda_{min}(P)$ and $\lambda_2 = \lambda_{max}(P)$, where $\lambda_{min}(P)$ and $\lambda_{max}(P)$ are the minimal and maximal eigenvalues of $P$, respectively. Then,

(i). $\left|\frac{\partial V}{\partial \eta_{\rho+1}}\right| \leq 2\lambda_{max}(P)\|\eta\|$, (ii) $W(\eta) \leq -\frac{V(\eta)}{\lambda_{max}(P)}$

(iii) $\|\eta\| \leq \sqrt{\frac{V(\eta)}{\lambda_{min}(P)}}$.

**Proof:** (i) Since $V(\eta) \leq \lambda_{max}(P)\|\eta\|^2$ and $\left|\frac{\partial V}{\partial \eta_{n+1}}\right| \leq \left\|\frac{\partial V(\eta)}{\partial \eta}\right\|$, then $\left|\frac{\partial V}{\partial \eta_{\rho+1}}\right| \leq 2\lambda_{max}(P)\|\eta\|$, (ii)Since $V(\eta) \leq \lambda_{max}(P)\|\eta\|^2 = \lambda_{max}(P)W(\eta)$, then, $-W(\eta) \leq \frac{V(\eta)}{\lambda_{max}(P)}$, and (iii) Since, $\lambda_{min}(P)\|\eta\|^2 \leq V(\eta)$, then, $\|\eta\| \leq \sqrt{\frac{V(\eta)}{\lambda_{min}(P)}}$ . □

**Theorem 1.** Given the chain of integrators system given in (6), and the ESO in (5). Then, under Assumptions (A1)-(A3), for any initial values,

$$\lim_{\substack{t \to \infty \\ \omega_0 \to \infty}} |\xi_i(t) - \hat{\xi}_i(t)| = 0 , i \in \{1,2. \ldots, \rho\}$$

$$\lim_{t \to \infty} |\hat{\xi}_{\rho+1} - (f(\xi_1, \xi_2, \ldots \xi_\rho, w, t) + (\beta(x) - b_0)u)| = 0$$

where $\xi_i(t)$, and $\hat{\xi}_i(t)$ denote the solutions of (6) and (5) respectively, $i \in \{1,2. \ldots, \rho+1\}$.

**Proof:** we make use of [14] to prove the convergence for the LESO. Set $e_i(t) = \xi_i(t) - \hat{\xi}_i(t)$, for $i \in \{1,2, \ldots, \rho+1\}$. Then subtracting (5) from (6), one gets

$$\begin{cases} \dot{\xi}_1(t) - \dot{\hat{\xi}}_1(t) = \xi_2(t) - \left(\hat{\xi}_2(t) + \beta_1\left(y(t) - \hat{\xi}_1(t)\right)\right) \\ \dot{\xi}_2(t) - \dot{\hat{\xi}}_2(t) = \xi_3(t) - \left(\hat{\xi}_3(t) + \beta_2(y(t) - \hat{\xi}_1(t))\right) \\ \quad \vdots \\ \dot{\xi}_\rho(t) - \dot{\hat{\xi}}_\rho(t) = \xi_{\rho+1}(t) + b_0 u(t) \\ \qquad - \left(\hat{\xi}_{\rho+1}(t) + b_0 u(t) + \beta_\rho\left(y(t) - \hat{\xi}_1(t)\right)\right) \\ \dot{\xi}_{\rho+1}(t) - \dot{\hat{\xi}}_{\rho+1}(t) = \Delta(t) - \beta_{\rho+1}(y(t) - \hat{\xi}_1(t)) \end{cases}$$

Direct computations show that the estimated error dynamics satisfy:

$$\begin{cases} \dot{e}_1(t) = e_2(t) - \beta_1 e(t) \\ \dot{e}_2(t) = e_3(t) - \beta_2 e(t) \\ \quad \vdots \\ \dot{e}_\rho(t) = e_{\rho+1}(t) - \beta_\rho e(t) \\ \dot{e}_{\rho+1}(t) = \Delta(t) - \beta_{\rho+1} e(t) \end{cases} \quad (9)$$

Let $\beta_i = a_i \omega_0^i$, where $a_i, i \in \{1,2. \ldots, \rho+1\}$ is the associated constant with each $\omega_0^i$, , and $\omega_0$ is the bandwidth. The final form with $\beta_i$ substituted in (9) is given as:

$$\begin{cases} \dot{e}_1(t) = e_2(t) - \omega_0 a_1 e_1(t) \\ \dot{e}_2(t) = e_3(t) - \omega_0^2 a_2 e_1(t) \\ \quad \vdots \\ \dot{e}_\rho(t) = e_{\rho+1}(t) - \omega_0^\rho a_\rho e_1(t) \\ \dot{e}_{\rho+1}(t) = \Delta(t) - \omega_0^{\rho+1} a_{\rho+1} e_1(t) \end{cases} \quad (10)$$

We assume $\eta_i(t) = \omega_0^{\rho+1-i} e_i(\frac{t}{\omega_0})$ , $i \in \{1,2, \ldots, \rho+1\}$ [16].

$$e_i\left(\frac{t}{\omega_0}\right) = \frac{1}{\omega_0^{\rho+1-i}} \eta_i(t) \quad (11)$$

Then time-scaled estimation error dynamics are expressed as:
$$\begin{cases} \frac{d\eta_1(t)}{dt} = \eta_2(t) - a_1\eta_1(t) \\ \frac{d\eta_2(t)}{dt} = \eta_3(t) - a_2\eta_1(t) \\ \vdots \\ \frac{d\eta_\rho(t)}{dt} = \eta_{\rho+1}(t) - a_n\eta_i(t) \\ \frac{d\eta_{\rho+1}(t)}{dt} = \frac{\Delta}{\omega_0} - a_{\rho+1}\eta_1(t) \end{cases} \quad (12)$$

Finding $\dot{V}$ (the differentiation of $V(\eta)$) w.r.t $t$ over $\eta$ (over the solution (12)) is accomplished in the following way

$$\dot{V}(\eta)|_{along\ (12)} = \sum_{i=1}^{\rho+1} \frac{\partial V(\eta)}{\partial \eta_i} \dot{\eta}_i(t)$$

$$= \sum_{i=1}^{\rho+1} \frac{\partial V(\eta)}{\eta_i}(\eta_{i+1}(t) - a_i\eta_1(t)) + \frac{\partial V(\eta)}{\partial \eta_{n+1}}\left(\frac{\Delta}{\omega_0} - a_{\rho+1}\eta_1(t)\right)$$

Then,
$$\dot{V}(\eta)|_{along\ (12)} = \sum_{i=1}^{\rho+1} \frac{\partial V(\eta)}{\eta_i}(\eta_{i+1}(t) - a_i\eta_1(t)) + \frac{\partial V(\eta)}{\partial \eta_{\rho+1}} \frac{\Delta}{\omega_0} - \frac{\partial V(\eta)}{\partial \eta_{\rho+1}} a_{\rho+1}\eta_1(t)$$

If the second inequality in Assumption (A3) is satisfied, then
$$\dot{V}(\eta)|_{along\ (12)} \leq -W(\eta) + \frac{\partial V(\eta)}{\partial \eta_{n+1}} \frac{\Delta}{\omega_0}$$

Given that the rate of change of the generalized disturbance is bounded (Assumption (A2)) and the results of Lemma 1, we get:
$$\dot{V}(\eta) \leq -\frac{V(\eta)}{\lambda_{max}(P)} + \frac{M}{\omega_0} 2\lambda_{max}(P) \frac{\sqrt{V(\eta)}}{\sqrt{\lambda_{min}(P)}} \quad (13)$$

knowning that $\frac{d}{dt}\sqrt{V(\eta)} = \frac{1}{2}\frac{1}{\sqrt{V(\eta)}}\dot{V}(\eta)$, then (13) is an ordinary 1st order differential equation (13), its solution can be found as,

$$\sqrt{V(\eta)} \leq \frac{2M\lambda^2_{max}(P)}{\omega_0\sqrt{\lambda_{min}(P)}}\left(1 - e^{-\frac{t}{2\lambda_{max}(P)}}\right) + \sqrt{V(\eta(0))}e^{-\frac{t}{2\lambda_{max}(P)}}$$

$$\|\eta(t)\| \leq \sqrt{\frac{1}{\lambda_{min}(P)}\left(\frac{2M\lambda^2_{max}(P)}{\omega_0\sqrt{\lambda_{min}(P)}}\left(1 - e^{-\frac{t}{2\lambda_{max}(P)}}\right) + \sqrt{V(\eta(0))}e^{-\frac{t}{2\lambda_{max}(P)}}\right)}$$

$$\|\eta(t)\| \leq \frac{2M\lambda^2_{max}(P)}{\omega_0\lambda_{min}(P)}\left(1 - e^{-\frac{t}{2\lambda_{max}(P)}}\right) + \sqrt{\frac{V(\eta(0))}{\lambda_{min}(P)}}e^{-\frac{t}{2\lambda_{max}(P)}}.$$ It follows from (11) that $|\xi_i(t) - \hat{\xi}_i(t)| = \frac{1}{\omega_{i0}^{\rho+1-i}}|\eta_i(\omega_0 t)| \Rightarrow |\xi_i(t) - \hat{\xi}_i(t)| \leq \frac{1}{\omega_0^{\rho+1-i}}\|\eta(t)\|$. Then, we get, $|\xi_i(t) - \hat{\xi}_i(t)|$
$\leq \frac{1}{\omega_0^{\rho+1-i}}\left(\frac{2M\lambda^2_{max}(P)}{\omega_0\lambda_{min}(P)}\left(1 - e^{-\frac{t}{2\lambda_{max}(P)}}\right) + \sqrt{\frac{V(\eta(0))}{\lambda_{min}(P)}}e^{-\frac{t}{2\lambda_{max}(P)}}\right)$.
Finally, $\lim_{\substack{t \to \infty \\ \omega_0 \to \infty}} |\xi_i(t) - \hat{\xi}_i(t)| = 0$ and
$\lim_{t \to \infty}|\hat{\xi}_{\rho+1} - (f(\xi_1, \xi_2, \ldots, \xi_\rho, w, t) + (\beta(x) - b_0)u)| = 0$   □

If the linearization control law (LCL) $u$ is selected as $u = v - \frac{\hat{\xi}_{\rho+1}}{b_0}$ and as result of theorem 1, then, the nonlinear system in (4) is reduced to a chain of $\rho$ integrators described as,
$$\xi_i = \xi_{i+1} \quad i \in \{1, 2, \ldots, \rho - 1\}, \quad \dot{\xi}_\rho = b_0 v \quad (14)$$

The main differences between IOFL and AIOFL are, for the AIOFL, there is no need to obtain the transformation of (3). The only required information is the relative degree of the system ($\rho$) for the nonlinear system to be linearized. While, for the IOFL, transformation (3) is the key step to linearize the system, it is based on exact mathematical cancelation of the nonlinear terms $\alpha(x)$ and $\beta(x)$, which requires knowledge of $\alpha$, $\beta$, and $T$. Furthermore, AIOFL in addition to linearizing the nonlinear system, it lumps the external disturbances, uncertainties, and unmodeled dynamics, into a single term for online and active estimation and cancelation later on.

## V. GUIDEWAY EXAMPLE

In this example, a SISO single-link flexible joint manipulator (SLFJM) offered by [15] is studied and shown in Fig. 2. The state-space representation of the SLFJM system in the form of the nonlinear system given in (1) is described as,
$$\begin{cases} \dot{x} = f(x) + bu + b_d\tau_d, \\ y = Cx. \end{cases}$$
Where $x = (x_1 \ x_2 \ x_3 \ x_4)^T = (\theta \ \alpha \ \dot{\theta} \ \dot{\alpha})^T \in \mathbb{R}^4$ is the plant state, $u \in \mathbb{R}$ the plant input, $\tau_d \in \mathbb{R}$ the exogenous disturbance, $y \in \mathbb{R}$ the plant output, and $f: \mathbb{R}^4 \to \mathbb{R}$. The components of $f, b, b_d, C$ are denoted, respectively, by

$$\begin{cases} f(x) = \begin{pmatrix} x_3 \\ x_4 \\ \frac{K_s}{J_h}x_2 - \frac{K_m^2 K_g^2}{R_m J_h}x_3 \\ -\frac{K_s}{J_h}x_2 - \frac{K_s}{J_l}x_2 + \frac{K_m^2 K_g^2}{R_m J_h}x_3 + \frac{mgh}{J_l}\sin(x_1 + x_2) \end{pmatrix} \\ b = \begin{pmatrix} 0 & 0 & \frac{K_m K_g}{R_m J_h} & -\frac{K_m K_g}{R_m J_h} \end{pmatrix}^T \\ b_d = \begin{pmatrix} 0 & 0 & \frac{1}{J_h} & -\frac{1}{J_h} \end{pmatrix}^T \\ C = (1 \ 1 \ 0 \ 0) \end{cases} \quad (15)$$

Where $K_s$ is the link stiffness, $J_h$ is the inertia of hub, $m$ is the link mass, $h$ is the height of hub, $K_m$ is the motor constant, $K_g$ is the gear ratio, $J_l$ is the load inertia, and $R_m$ is the motor resistance. The values of the coefficients for SLFJM are [15]: $K_s = 1.61$, $J_h = 0.0021$, $m = 0.403$, $g = -9.81$, $h = 0.06$, $K_m = 0.00767$, $K_g = 70$, $J_l = 0.0059$, and $R_m = 2.6$. Applying the Lie derivative on equation (5), we get the following set of equations [16]:
$$L_g h(x) = 0, L_g L_f^1 h(x) = 0,$$
$$L_g L_f^2 h(x) = 0, L_g L_f^3 h(x) \neq 0$$

It can be noticed that the SLFJM system in (15) satisfies (2); consequently, the relative order of SLFJM is 4, i.e., $\rho = 4$ [16]. Two types of ESOs are tested in this paper. The conventional ADRC is the combination of the Linear ESO(LESO) given by (16), NLSEF was given by (17), and TD given by (18). According to [13], [14], a LESO observer can be designed as:

$$\begin{cases} \dot{\hat{\xi}}_1 = \hat{\xi}_2 + \beta_1(y - \hat{\xi}_1), \dot{\hat{\xi}}_2 = \hat{\xi}_3 + \beta_2(y - \hat{\xi}_1) \\ \dot{\hat{\xi}}_3 = \hat{\xi}_4 + \beta_3(y - \hat{\xi}_1), \dot{\hat{\xi}}_4 = \hat{\xi}_5 + \beta_4(y - \hat{\xi}_1) + b_0 u \\ \dot{\hat{\xi}}_5 = \beta_5(y - \hat{\xi}_1) \end{cases} \quad (16)$$

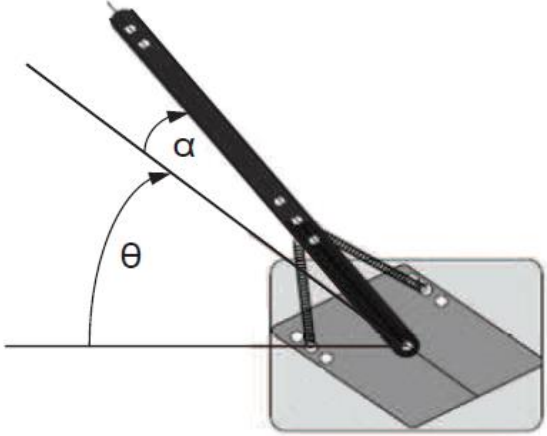

Fig. 2 Definition of generalized coordinates for the SLFJM

Where $\hat{\xi} = (\hat{\xi}_1 \ \hat{\xi}_2 \ \hat{\xi}_3 \ \hat{\xi}_4 \ \hat{\xi}_5)^T$ is the observer's state vector, and $\beta = (\beta_1 \ \beta_2 \ \beta_3 \ \beta_4 \ \beta_5)^T = (\alpha_1\omega_0 \ \alpha_2\omega_0^2 \ \alpha_3\omega_0^3 \ \alpha_4\omega_0^4 \ \alpha_5\omega_0^5)^T$ is the observer gain vector. The control law for the AIOFL is given as, $u = v - \frac{\hat{\xi}_5}{b_0}$. Where $v = fal(e_1. \alpha_1. \delta_1) + fal(e_2. \alpha_2. \delta_2)$, with

$$fal(e.\alpha.\delta) = \begin{cases} \frac{e}{\delta^{\alpha-1}} & |e| \leq \delta \\ |e|^\alpha sgn(e) & |e| > \delta \end{cases} \quad (17)$$

where $e = (e_1 \ e_2)^T$ is the tracking error vector which can be defined as $e_i = r_i - \hat{\xi}_i$ with $i = 1, 2$. The nonlinear second order differentiator is given as [11]:

$$\begin{cases} \dot{r}_1 = r_2 \\ \dot{r}_2 = -R\, sign(r_1 - r(t) + \frac{r_2|r_2|}{2r}) \end{cases} \quad (18)$$

Where $r_1$ is tracking signal of the input $r$, and $r_2$ tracking signal of the derivative of the input $r$. To speed up or slow down the system during the transient, the coefficient $R$ is adapted according to this, it is an application dependent.

The Improved ADRC(IADRC) has been designed in our previous works [12], [17], [18] and tested on the differential drive mobile robot model [19]. It is structured from the improved nonlinear extended state observer (INLESO) given by (19) [12], improved NLSEF (INLSEF) given by (20) [17], and improved TD(ITD) given by (21)[18]. The INLESO is the second type of observers used in this numerical simulations. The INESO is described as:

$$\begin{cases} \dot{\hat{\xi}}_1 = \hat{\xi}_2 + \beta_1 g(y - \hat{\xi}_1), \dot{\hat{\xi}}_2 = \hat{\xi}_3 + \beta_2 g(y - \hat{\xi}_1) \\ \dot{\hat{\xi}}_3 = \hat{\xi}_4 + \beta_3 g(y - \hat{\xi}_1), \dot{\hat{\xi}}_4 = \hat{\xi}_5 + \beta_4 g(y - \hat{\xi}_1) + b_0 u \\ \dot{\hat{\xi}}_5 = \beta_5 g(y - \hat{\xi}_1) \end{cases} \quad (19)$$

Where $g(e) = k_\alpha |e|^\alpha sign(e) + k_\beta |e|^\beta e$, the two vectors $\xi$ and $\beta$ are defined previously as in LESO case. The control law for the AIOFL is defined as $u = v - \frac{\hat{\xi}_5}{b_0}$. Where $v$ is given in our previous work as[17]:

$$v = \delta \tanh\left(\frac{v_1 + v_2}{\delta}\right) \quad (20)$$

$$v_1 = \left(k_{11} + \frac{k_{12}}{1 + \exp(\mu_1 e_1^2)}\right) |e_1|^{\alpha_1} sign(e_1)$$

$$v_2 = \left(k_{21} + \frac{k_{22}}{1 + \exp(\mu_2 e_2^2)}\right) |e_2|^{\alpha_2} sign(e_2)$$

The tracking differentiator associated with INLESO is described as [12]:

$$\begin{cases} \dot{r}_1 = r_2 \\ \dot{r}_2 = -\rho^2 \tanh\left(\frac{br_1 - (1-a)r}{c}\right) - \rho r_2 \end{cases} \quad (21)$$

Where the coefficients $a, b, c,$ and $\rho$ are suitable design factors, where $0 < a < 1, b > 0, c > 0,$ and $\rho > 0$. The AIOFL based on the classical LESO and INLESO is applied on the SLFJM given in (5). An Objective Performance Index (OPI) is proposed to evaluate the performance of the LESO and the INESO observers, which is represented as:

$$\text{OPI} = w_1 \frac{\text{ITAE}}{N_1} + w_2 \frac{\text{ISU}}{N_2} + w_3 \frac{\text{IAU}}{N_3} \quad (22)$$

Where $\text{ITAE} = \int_0^{t_f} t|y - r|dt$ is the integration of the time absolute error for the output signal, $\text{ISU} = \int_0^{t_f} u_o^2 \, dt$ is the integration of square of the control signal, and $\text{IAU} = \int_0^{t_f} |u_o| \, dt$ is the integration of the absolute of the control signal. The weights must satisfy $w_1 + w_2 + w_3 = 1$, are defined as the relative emphasis of one objective as compared to the other. The values of $w_1, w_2,$ and $w_3$ are chosen to increase the pressure on selected objective function. The $N_1, N_2,$ and $N_3$ ar included in the performance index to insure that the individual objectives have comparable values, and are treated equally likely by the tuning algorithm. Because, if a certain objective is of very high value, while the second one has very low value, then the tuning algorithm will pay much consideration to the highest one and leave the other with little reflection on the system. The tuning process of both observers is achieved using GA under MATLAB environment with $w_1 = 0.6, w_2 = 0.2, w_3 = 0.6, N_1 = 10, N_2 = 2, N_3 = 2.7,$ and $t_f = 6$ sec. Based on this, the tuned parameters values for the LESO in (16) and the associated control law in (17) are: $\omega_0 = 513.8283, \alpha_1 = 8.772, \alpha_2 = 0.1946, \alpha_3 = 0.7384, \alpha_4 = 9.6881 \times 10^{-3}, \alpha_5 = 2.2651 \times 10^{-6}, b_0 = 22.771, R = 2408.6918, \delta_1 = 16.6108, \delta_2 = 14.6238, \alpha_1 = 0.3804,$ and $\alpha_2 = 0.4583$. The values of tuned parameters for INESO and associated control law: $k_{11} = 1.7741, k_{12} = 1.2147, k_{21} = 0.00115, k_{22} = 0.3312, \delta = 3.3900, \mu_1 = 3.8297, \mu_2 = 10.9415, \alpha 1 = 0.8244, \alpha 2 = 1.8079, a = 0.9153, b = 8.7141, c = 0.0813, \rho = 22.89333, \omega_0 = 104.6131, \alpha 1 = 0.1364, \alpha 2 = 0.6691, \alpha 3 = 0.6893, \alpha 4 = 0.0155, \alpha 5 = 14.3801 \times 10^{-6}, b_0 = 8.74500, \alpha = 0.6906, \beta = 0.1880,$

$k_\alpha = 0.3682$, $k_\beta = 0.1290$. It is worthy to note that in the AIOFL, the nonlinear system is linearized by either LESO or INESO and represented by a chain of integrators given by (14). In this case, the ESO (linear or nonlinear) will estimate the states of the chain of integrators up to the relative degree of the nonlinear system. With this arrangement, the higher order estimated states represent signals with higher derivative degrees, they contain high frequency components which in turn increase the control signal activity and leading to the chattering phenomena. Based on the above reasoning, only the first two estimated states ($\hat{\xi}_1$ and $\hat{\xi}_2$) are fedback to either NLSEF or INLSEF in the simulations. On contrast to the conventional ADRC, where the entire estimated states of the system (except the augmented state) are provided for feedback to the NLSEF. In our case, with the first two estimated states, it was sufficient to produce the individual control laws ($v_1$ and $v_2$) which in turn produced the required control law ($v$). With this scenario, eliminating the states from the feedback that do not affect on the performance of the system will reduce the number of the parameters of both the NLSEF controller and the TD. We expect that the total energy required for the controller to produce the control law ($v$) will be reduced. Runge-Kutta ODE45 solver in MATLAB environment has been used for the numerical simulations of the continuous models. A sinusoidal signal with frequency 2 rad/sec and amplitude of 45 has been chosen as a reference input. The simulation time is selected to be 20 sec. The results of the numerical simulation for the AIOFL of the two test cases are shown in Fig 3. The results are collected based on evaluating two indices listed in tables I. Where ITAE = $\int_0^{20} t|y - r|dt$ is the integration of the time absolute error for the output signal, and ISU = $\int_0^{20} u^2\, dt$ is the integration of square of the control signal. The simulations show that the ISU index, which represents the energy delivered to the SJFLM motor, has been decreased by 23.82% and a noticeable improvement in the transient response (ITAE is reduced by 23.7%).

Table I The results of the numerical simulation

| AIOFL structure | ITAE | ISU |
|---|---|---|
| LESO | 126.120273 | 7.982831 |
| INLESO | 96.225965 | 6.080829 |

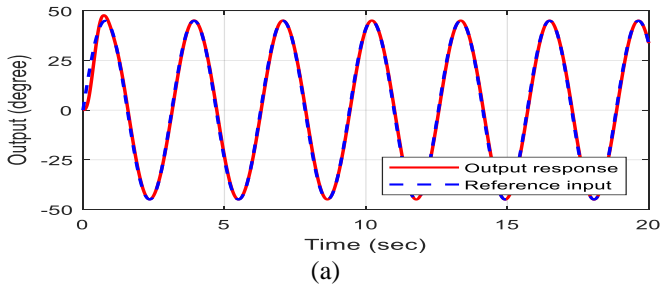

(a)

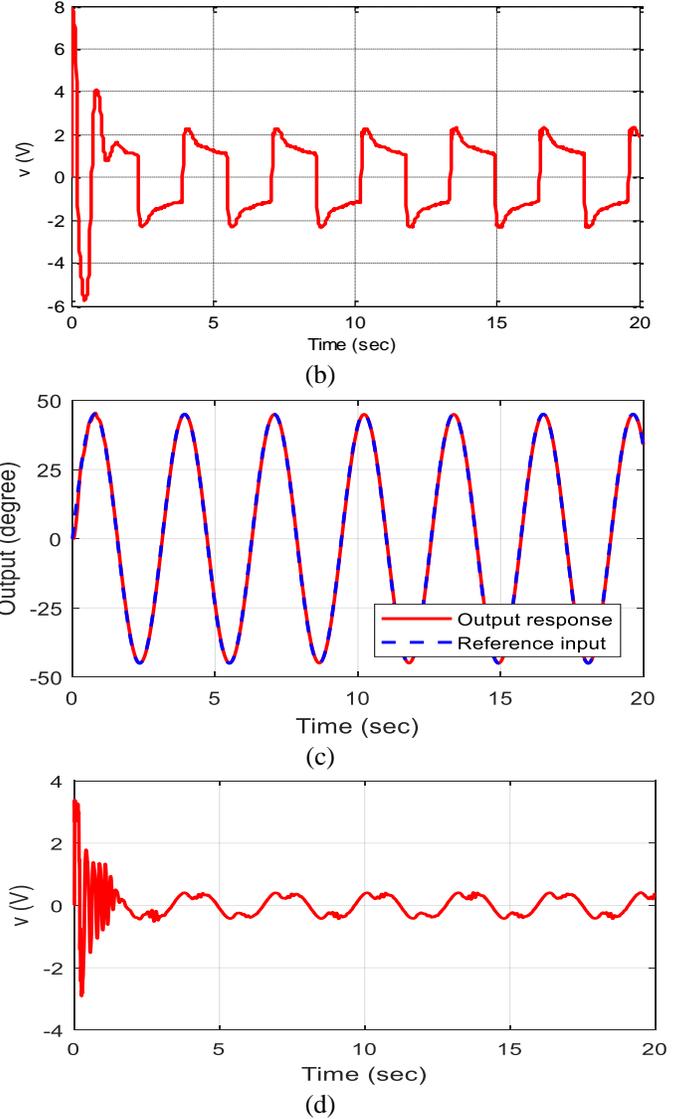

Fig. 3 The curves of the numerical simulations, (a)The output response of the SLFJM for LESO (b)The control signal for LESO (c) The output response for INLESO (d)The control signal for INLESO

A second simulation scenario considered in this work is included the presence of an exogenous disturbance $\tau_d$ of type step at t = 10 sec with amplitude of 0.5 $N.m$ and an increase 40% in the load inertia. The results of the numerical simulation are shown in Fig. 4. The numerical results of the two performance indices of the second scenario are listed in tables II. As shown in Table II, the ITAE significantly reduced for the INLESO case. This improvement in the transient response, which is reflected by the value of ITAE goes along with insignificant increase in the delivered energy to the actuation.

Table II The results of the numerical simulation

| AIOFL structure | ITAE | ISU |
|---|---|---|
| LESO | 1309.213956 | 58.214189 |
| INLESO | 298.143303 | 69.471044 |

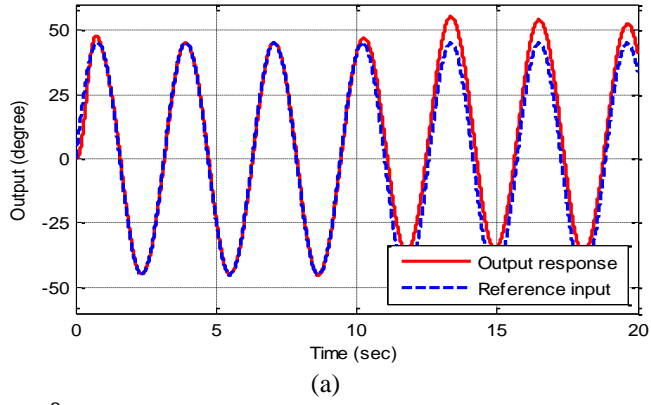

(a)

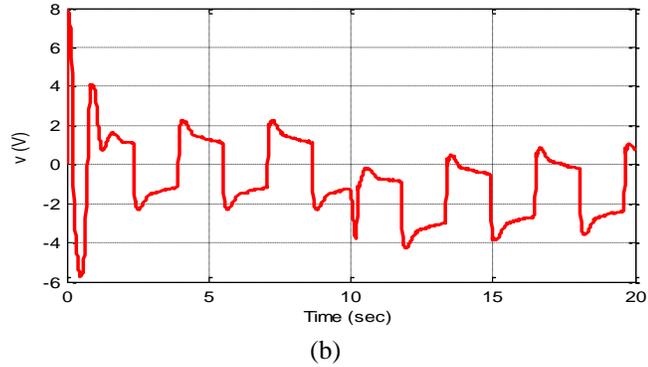

(b)

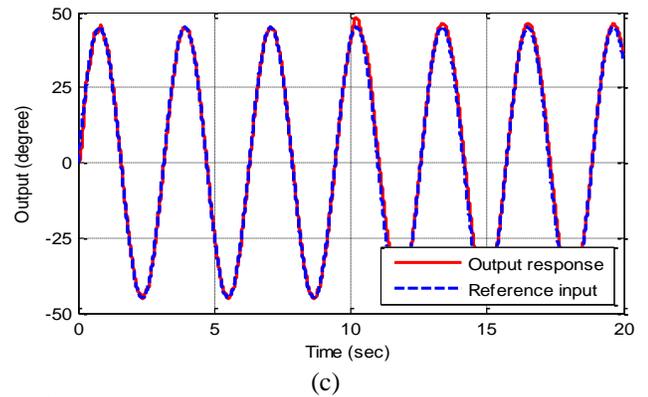

(c)

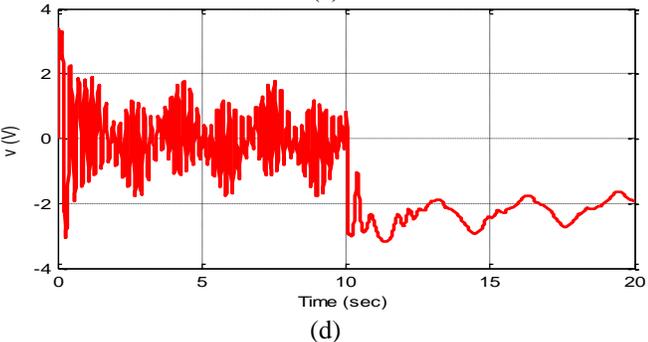

(d)

Fig. 4 The curves of the numerical simulations for the second scenario, (a)The output response of the SLFJM for LESO (b)The control signal for LESO (c) The output response for INLESO (d)The control signal for INLESO

The final scenario that has been done in this work is testing the immunity of the system against noise. A Gaussian measurement noise at the output is considered, the variance and mean of the Gaussian noise are 0.0001 and 0, respectively. To actively counteract the effect of the noise, both ESOs are re-tuned again using GA under the existence of noise based on the OPI defined in (22). The new tuned parameters of the LESO are $\omega_0 = 851.0106$, $\alpha_1 = 5.40326$, $\alpha_2 = 0.2871$, $\alpha_3 = 0.7644$, $\alpha_4 = 0.01$, $\alpha_5 = 1.22x10^{-6}$, and $b_0 = 33.7432$. While the new tuned parameters of the INLESO are $\omega_0 = 121.020$, $\alpha_1 = 0.205$, $\alpha_2 = 0.6$, $\alpha_3 = 0.42$, $\alpha_4 = 0.0232$, $\alpha_5 = 7.19x10^{-6}$, and $b_0 = 9.7$. The results of the numerical simulation are shown in Fig. 5. The numerical results of the two performance indices of the second scenario are listed in tables III. As shown in Table III, both of the ITAE and ISU are reduced significantly using the INESO. This improvement in the transient response and reduced the delivered energy is noticeable shown in Fig. 5.

Table III The results of the numerical simulation

| AIOFL structure | ITAE | ISU |
|---|---|---|
| LESO | 332.443873 | 799.520367 |
| INLESO | 102.578228 | 19.959797 |

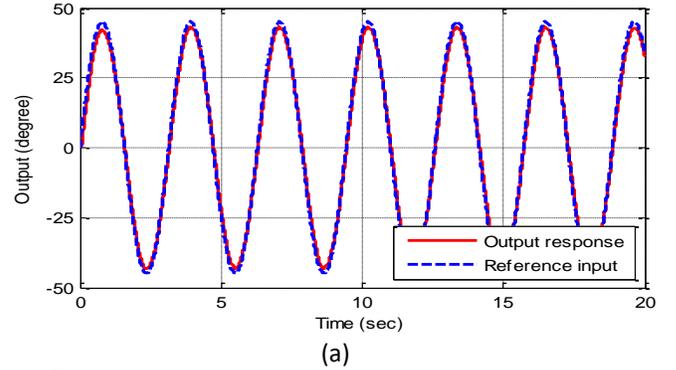

(a)

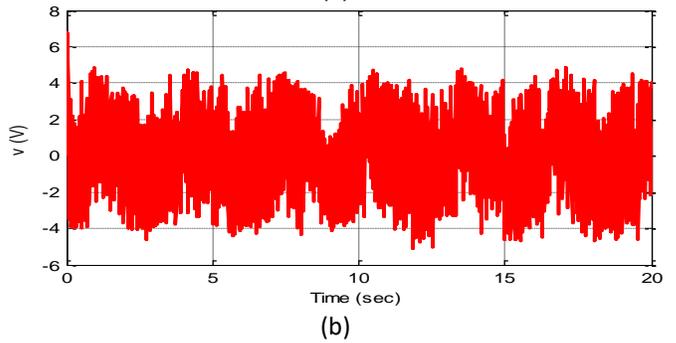

(b)

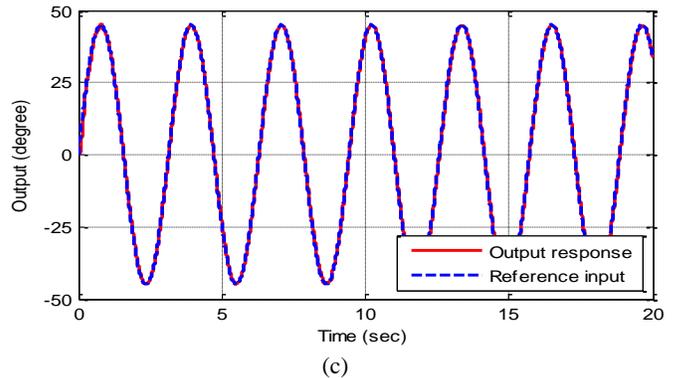

(c)

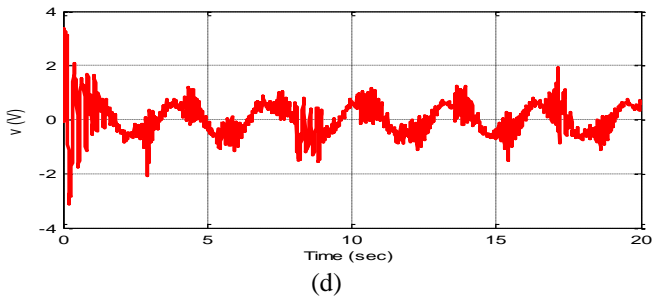

(d)

Fig. 5 The curves of the numerical simulations for the third scenario, (a)The output response of the SLFJM for LESO (b)The control signal for LESO (c) The output response for INLESO (d)The control signal for INLESO

VI. CONCLUSIONS

This paper addressed the problem of AIOFL for general uncertain nonlinear system subjected to external disturbances. It differs from the traditional IOFL, which assumes a nominal nonlinear system to work on it. The AIOFL has been implemented by extended state observer, which transforms the nonlinear uncertain system into a chain of integrators. The key point of the proposed methods is that it requires only the relative degree of the nonlinear uncertain system. It can be concluded that the proposed AIOFL scenario transform any nonlinear system into a linear one and excellently estimate and cancels the generalized disturbance. The estimation error is inversely proportional to the bandwidth of the nonlinear system and the proposed AIOFL is asymptotically stable using the designed ESOs. While both versions of the designed ESOs present good tracking, the NLESO exhibits better performance than its linear one and provides the actuator with a more stable control signal, it has less fluctuation with little amplitude.